\documentclass[aps,twocolumn,nofootinbib,superscriptaddress,preprintnumbers,pra]{revtex4-1}

\usepackage{float}
\usepackage{arydshln}
\usepackage{colortbl}
\usepackage{amsmath,bbm,blkarray,bm,amssymb}
\usepackage{dsfont}
\usepackage{hyperref}
\usepackage{graphicx}
\usepackage{enumitem}
\usepackage{arydshln}
\usepackage{bbold}
\usepackage{soul}
\usepackage{slashed}
\usepackage{yfonts}
\usepackage{tikz}
\usepackage{changepage}
\usepackage{makecell}
\usepackage{tikz-feynman}
\usepackage{subcaption}
\usepackage{color}
\usepackage[T1]{fontenc}
\usepackage{lipsum}
\usepackage{slashed}
\usepackage{mathtools}
\usepackage{physics}
\usepackage{booktabs}
\usepackage{multirow}
\usepackage[vcentermath]{youngtab}

\makeatletter
\g@addto@macro\bfseries{\boldmath}
\makeatother

\newcommand{\be} {\begin{equation}}
\newcommand{\ee} {\end{equation}}
\newcommand{\bea} {\begin{eqnarray}}
\newcommand{\eea} {\end{eqnarray}}

\renewcommand{\Re}{{\rm Re}}

\newcommand{\Op}[3]{ #1^{#2}_{#3}}
\newcommand{\cop}[3]{\widetilde #1^{#2}_{#3}}
\newcommand{\1}{\mathbb{1}}
\newcommand{\D}[2]{\mathcal{D}_{#1}^{#2}}

\begin{document}

\title{Dipole operators in Fierz identities}

\author{Jason Aebischer}
\author{Marko Pesut}
\author{Zachary Polonsky}

\affiliation{Physik-Institut, Universit\"at Z\"urich, CH-8057 Z\"urich, Switzerland}

\begin{abstract}
\vspace{5mm}
We study the contribution from dipole operators to one-loop Fierz identities and provide the resulting QCD and QED shifts to the tree-level relations for all four-fermion operators. The results simplify one-loop basis changes as well as matching computations and allow one to consistently eliminate operators from an operator basis which give rise to complications, e.g. traces involving $\gamma_5$.
\vspace{3mm}
\end{abstract}

\maketitle
\allowdisplaybreaks

\section{Introduction}
\vspace{-0.5cm}
Fierz identities relate tensor products of Dirac structures in $D=4$ space-time dimensions
\cite{Fierz:1939zz}. When using dimensional regularization, space-time is continued to
$D=4-2\varepsilon$ dimensions and the breaking of Fierz identities is handled by introducing
evanescent operators \cite{Buras:1989xd,Dugan:1990df,Herrlich:1994kh}. Such evanescent structures, when
inserted into divergent loop diagrams lead to finite contributions, which can be interpreted as one-loop
shifts to the regular Fierz transformations. In Ref.~\cite{Ciuchini:1997bw}, such shifts are avoided by fixing
the renormalization scheme (and therefore operator basis) to preserve tree-level Fierz relations; the Regularization Independent (RI) scheme.
While this scheme avoids the issue of shifted Fierz identities, it has the downside that the now-fixed operator basis
may not be the most convenient choice for the calculation at hand. For example, the
basis may include operators that result in traces featuring $\gamma_5$ when computing loop-level diagrams that may
be avoided in a different basis.

A different approach is to fix the operator basis which is most convenient to the calculation at hand and directly account for
the modification of Fierz identities. This strategy has been successfully applied in the case of $\Delta F=1$ processes \cite{Aebischer:2022tvz,Aebischer:2021raf,Aebischer:2021hws} and $\Delta F=2$ \cite{Aebischer:2022anv}, as well as in matching computations involving Leptoquarks (LQs) \cite{Aebischer:2018acj}. Furthermore, when performing two-loop computations, such shifts are
necessary to take into account since they describe the mixing of evanescent operators into the physical sector
\cite{Herrlich:1994kh}. The shifts for all four-fermion operators, together with the
corresponding renormalization constants, have been computed recently in \cite{Aebischer:2022aze} in the
generalized BMU scheme \cite{Buras:2000if}. The effects of dipole operators on the one-loop QCD and QED shifts were not discussed in~\cite{Aebischer:2022aze}, since mass effects were neglected. In Ref.~\cite{Fuentes-Martin:2022vvu}, dipole contributions to SMEFT basis changes were calculated using path-integral techniques. In this letter we give all one-loop shifts that result from electric and chromomagnetic dipole operators and therefore complete the findings in \cite{Aebischer:2022aze} up to one-loop order in QCD and QED for effective four-fermion operators up to mass-dimension-six.

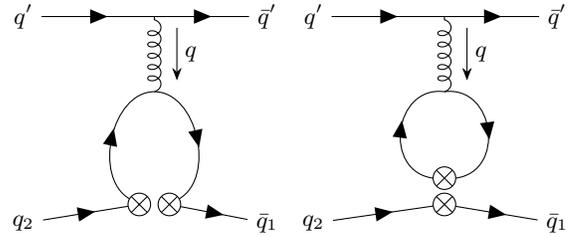
\begin{figure}[b]
	\centering
	\begin{tikzpicture}
	\begin{feynman}
		\vertex (i4) {\(\bar{q}'\)};
		\vertex[left=1.5cm of i4] (a);
		\vertex[left=1.5cm of a] (i3) {\(q'\)};
		\vertex[below=1cm of a] (b);
		\vertex[below=1.5cm of b] (fake);
		\node[right=0.05cm of fake, crossed dot] (j2);
		\node[left=0.05cm of fake, crossed dot] (j1);
		\vertex[below=2.75cm of i3] (i1) {\(q_2\)};
		\vertex[below=2.75cm of i4] (i2) {\(\bar{q}_1\)};
		\diagram*{
		(i1) --[fermion] (j1) --[fermion, in = 180, out = 150] (b)
		--[fermion, in = 30, out = 0] (j2) --[fermion] (i2),
		(b) --[gluon, reversed momentum' = \(q\)] (a),
		(i3) --[fermion] (a) --[fermion] (i4)
		};
	\end{feynman}
	\end{tikzpicture}
	\begin{tikzpicture}
	\begin{feynman}
		\vertex (i4) {\(\bar{q}'\)};
		\vertex[left=1.5cm of i4] (a);
		\vertex[left=1.5cm of a] (i3) {\(q'\)};
		\vertex[below=1cm of a] (b);
		\node[below=1cm of b, crossed dot] (c);
		\node[below=0.35cm of c, crossed dot] (d);
		\vertex[below=2.75cm of i3] (i1) {\(q_2\)};
		\vertex[below=2.75cm of i4] (i2) {\(\bar{q}_1\)};
		\diagram*{
		(i1) --[fermion] (d) --[fermion] (i2),
		(c) --[fermion, half left, looseness=1.5] (b) --[fermion, half left, looseness=1.5] (c),
		(b) --[gluon, reversed momentum' = \(q\)] (a),
		(i3) --[fermion] (a) --[fermion] (i4)
		};
	\end{feynman}
	\end{tikzpicture}
	\caption{Two topologies of penguin diagrams with four-fermion operator insertions: open (left)
	and closed (right). Here, the fermion lines follow spinor indices.}
	\label{fig:topologies}
\end{figure}

\section{Procedure}

In order to obtain one-loop shifts to tree-level Fierz relations, one considers insertions of four-fermion operators into one-loop diagrams. The
tree-level transformation from an operator basis, $\{\mathcal{O}_i\}$ to its Fierz-conjugated counterpart, $\{\tilde{\mathcal{O}}_i\}$ is given
by
\begin{equation}\label{eq:tree_fierz}
	\big<\tilde{\mathcal{O}}_i\big>^{(0)} = \mathcal{F}_{ij}\big<\mathcal{O}_j\big>^{(0)}\,.
\end{equation}
Denoting the one-loop amplitudes with $\mathcal{O}$ and $\tilde{\mathcal{O}}$ operator insertions
\begin{equation}
	\begin{split}
		\big<\mathcal{O}_i\big>^{(1)} =& \Big(\delta_{ij} + r^{(1)}_{ij}\Big)\big<\mathcal{O}_j\big>^{(0)}\,, \\[0.5em]
		\big<\tilde{\mathcal{O}}_i\big>^{(1)} =& \Big(\delta_{ij} + \tilde{r}^{(1)}_{ij}\Big)\big<\tilde{\mathcal{O}}_j\big>^{(0)}\,,
	\end{split}
\end{equation}
respectively, the corresponding one-loop-corrected transformation is given by
\begin{equation}
	\Big(\delta_{ij} + \tilde{r}_{ij}^{(1)}\Big)\big<\tilde{\mathcal{O}}_j\big>^{(0)} = 
	\Big(\mathcal{F}_{ij} + \mathcal{F}_{ik}r^{(1)}_{kj} + \Delta_{ij}\Big)\big<\mathcal{O}_j\big>^{(0)}\,,
\end{equation}
where $\Delta$ is the one-loop correction to the tree-level Fierz transformation. Using Eq.~\ref{eq:tree_fierz} yields
\begin{equation}
	\Delta_{ij} = \tilde{r}^{(1)}_{ik}\mathcal{F}_{kj} - \mathcal{F}_{ik}r^{(1)}_{kj}\,.
\end{equation}

In the case of dipole operators, only penguin diagrams need to be considered in the computation of $r^{(1)}$ and $\tilde{r}^{(1)}$, since genuine vertex corrections do not produce dipole structures. Furthermore, one-loop shifts arise from $\mathcal{O}(\epsilon)$ terms in $D$-dimensional Dirac algebra multiplied by $1/\epsilon$ poles from loops, so only the divergent parts of the loop integrals need to be calculated.
Therefore, considering the general operators
\begin{align}
  Q_o &= (\overline q_1 \Gamma_1^o q_3)(\overline q_3 \Gamma_2^o q_2)\,,  \\
  Q_c &= (\overline q_1^\alpha \Gamma_1^c q_2^\beta)(\overline q_3^\beta \Gamma_2^c q_3^\alpha)\,,
\end{align}
together with our convention for the covariant derivative of a quark field
\begin{equation}
	D_\mu q = \left[\partial_\mu +ieQ_q A_\mu+ig_sT^A G_\mu^A \right] q\,,
\end{equation}
we consider the following master formulae for the pole structure of the open and closed penguin diagrams shown in Fig.~\ref{fig:topologies}:\footnote{
Here, we have given the results for diagrams where the internal line is a gluon. When it is a photon,
the $SU(3)$ generators are replaced by the respective electric charges of the fermions and the strong coupling constant is replaced by the electric one.}
\begin{align}
  P^o_{\rm div} &= i\frac{\alpha_s}{4\pi}\frac{1}{q^2}(\overline q'\gamma^\mu T^A q')\times \notag\\
  &(\overline q_1 \Gamma_1^o\left[\frac{D}{12}q^2+m_{q_3}^2(2-\frac{D}{2})+m_{q_3}\slashed{q}\right]
	\gamma_\mu \Gamma_2^oT^Aq_2)\frac{1}{\epsilon}\,, \\
  P^c_{\rm div} &= -i\frac{\alpha_s}{4\pi}\frac{1}{q^2}(\overline q'\gamma^\mu T^A q')
	(\overline q_1\Gamma^c_1 T^Aq_2)\times \notag\\\label{eq:closed}
   &\left[{\rm Tr}\left[\gamma_\mu\Gamma_2^c\right]\left(\frac{D}{12}q^2+m_{q_3}^2
	(2-\frac{D}{2})\right)+m_{q_3}{\rm Tr}\left[\slashed{q}\gamma_\mu\Gamma_2^c\right]\right]
	\frac{1}{\epsilon}\,.
\end{align}
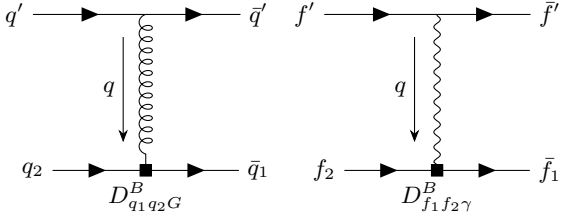
\begin{figure}[t]
        \centering
        \begin{tikzpicture}
        \begin{feynman}
                \vertex (i4) {\(\bar{q}'\)};
                \vertex[left=1.5cm of i4] (a);
                \vertex[left=1.5cm of a] (i3) {\(q'\)};
                \node[below=2cm of a, square dot] (b);
		\node[below=0.35cm of b] (label) {\(D_{q_1q_2G}^B\)};
                \vertex[left=1.5cm of b] (i1) {\(q_2\)};
                \vertex[right=1.5cm of b] (i2) {\(\bar{q}_1\)};
                \diagram*{
			(i1) --[fermion] (b) --[fermion] (i2),
			(a) --[gluon, momentum' = \(q\)] (b),
			(i3) --[fermion] (a) --[fermion] (i4)
                };
        \end{feynman}
        \end{tikzpicture}
        \begin{tikzpicture}
        \begin{feynman}
                \vertex (i4) {\(\bar{f}'\)};
                \vertex[left=1.5cm of i4] (a);
                \vertex[left=1.5cm of a] (i3) {\(f'\)};
                \node[below=2cm of a, square dot] (b);
		\node[below=0.35cm of b] (label) {\(D_{f_1f_2\gamma}^B\)};
                \vertex[left=1.5cm of b] (i1) {\(f_2\)};
                \vertex[right=1.5cm of b] (i2) {\(\bar{f}_1\)};
                \diagram*{
			(i1) --[fermion] (b) --[fermion] (i2),
			(a) --[photon, momentum' = \(q\)] (b),
			(i3) --[fermion] (a) --[fermion] (i4)
                };
        \end{feynman}
        \end{tikzpicture}
	\caption{Tree-level Feynman diagrams contributing to $\D{q_1q_2G}{B}$ (left) and
	$\D{f_1f_2\gamma}{B}$ (right), where the square dot represents a dipole operator insertion
	and $B = R,L$ denotes the chirality of the dipole fermion line.}
        \label{fig:dipoles}
\end{figure}
The resulting shifts are then obtained by considering all possible operator insertions into the above formulae and projecting the difference of the two onto the dipole operators. In the next section we will present the results for all possible four-fermion operators. We report the results in the basis used in \cite{Aebischer:2022aze}, for which we adopt the following notation for four-quark operators
\begin{align}\label{eq:V4q}
  V^{AB}_{q_1q_2q_3q_4} &= (\overline q_1^\alpha\gamma_\mu P_A q_2^\alpha)(\overline q_3^\beta\gamma^\mu P_B q_4^\beta)\,,\\
  S^{AB}_{q_1q_2q_3q_4} &= (\overline q_1^\alpha P_A q_2^\alpha)(\overline q_3^\beta P_B q_4^\beta)\,,\\
  T^{AB}_{q_1q_2q_3q_4} &= (\overline q_1^\alpha\sigma_{\mu\nu} P_A q_2^\alpha)(\overline q_3^\beta\sigma^{\mu\nu} P_B q_4^\beta)\,,\\
  {\widetilde V}^{AB}_{q_1q_2q_3q_4} &= (\overline q_1^\alpha\gamma_\mu P_A q_2^\beta)(\overline q_3^\beta\gamma^\mu P_B q_4^\alpha)\,, \\
  {\widetilde S}^{AB}_{q_1q_2q_3q_4} &= (\overline q_1^\alpha P_A q_2^\beta)(\overline q_3^\beta P_B q_4^\alpha)\,, \\
  {\widetilde T}^{AB}_{q_1q_2q_3q_4} &= (\overline q_1^\alpha \sigma_{\mu\nu}P_A q_2^\beta)(\overline q_3^\beta \sigma^{\mu\nu}P_B q_4^\alpha)\,,
\end{align}
the semi-leptonic operators
\begin{align}
  V^{AB}_{q_1q_2\ell_1\ell_2} &= (\overline q_1^\alpha\gamma_\mu P_A q_2^\alpha)(\overline \ell_1\gamma^\mu P_B \ell_2)\,,\\
  S^{AB}_{q_1q_2\ell_1\ell_2} &= (\overline q_1^\alpha P_A q_2^\alpha)(\overline \ell_1 P_B \ell_2)\,,\\
  T^{AB}_{q_1q_2\ell_1\ell_2} &= (\overline q_1^\alpha\sigma_{\mu\nu} P_A q_2^\alpha)(\overline \ell_1\sigma^{\mu\nu} P_B \ell_2)\,,\\
  V^{AB}_{q_1\ell_2\ell_1q_2} &= (\overline q_1^\alpha\gamma_\mu P_A \ell_2)(\overline \ell_1\gamma^\mu P_B q_2^\alpha)\,, \\
  S^{AB}_{q_1\ell_2\ell_1q_2} &= (\overline q_1^\alpha P_A \ell_2)(\overline \ell_1 P_B q_2^\alpha)\,, \\
  T^{AB}_{q_1\ell_2\ell_1q_2} &= (\overline q_1^\alpha \sigma_{\mu\nu}P_A \ell_2)(\overline \ell_1 \sigma^{\mu\nu}P_B q_2^\alpha)\,,
\end{align}
and the four-lepton operators
\begin{align}\label{eq:V}
  V^{AB}_{\ell_1\ell_2\ell_3\ell_4} &= (\overline \ell_1\gamma_\mu P_A \ell_2)(\overline \ell_3\gamma_\mu P_B \ell_4)\,, \\
  S^{AB}_{\ell_1\ell_2\ell_3\ell_4} & = (\overline \ell_1 P_A \ell_2)(\overline \ell_3 P_B \ell_4)\,, \\
  T^{AB}_{\ell_1\ell_2\ell_3\ell_4} & = (\overline \ell_1 \sigma_{\mu\nu}P_A \ell_2)(\overline \ell_3 \sigma^{\mu\nu}P_B \ell_4)\,,\label{eq:T4l}
\end{align}
where $P_A,P_B = P_{R/L} = \frac{1}{2}(\1\pm\gamma_5)$ are the right/left fermion projection operators and $\alpha,\beta$ are color indices.

In addition, we introduce the chromo- and electromagnetic dipole operators
\begin{align}
  D_{q_1q_2 G}^B &= \frac{1}{g_s}m_q (\overline q_1 \sigma^{\mu\nu}P_B T^A q_2)G^A_{\mu\nu}\,,\\
  D_{f_1f_2 \gamma}^B &= \frac{1}{e}m_f (\overline f_1 \sigma^{\mu\nu}P_B  f_2)F_{\mu\nu}\,.
\end{align}
where $f\in\{q,\ell\}$.

Using this notation we find for the tree-level amplitudes with $D_{q_1q_2 G}^L$ and $D_{f_1f_2 \gamma}^L$insertions with incoming momentum $q$, depicted in Fig.~\ref{fig:dipoles}:

\begin{align}
  -i\frac{2m_q}{q^2}(\overline q'\gamma^\mu T^Aq')(\overline q_1\slashed{q}\gamma_\mu P_LT^Aq_2)&\equiv
	\D{q_1q_2G}{L}\,, \\
	-i\frac{2m_f}{q^2}Q^\prime(\overline f'\gamma^\mu f')(\overline f_1\slashed{q}\gamma_\mu P_Lf_2)&\equiv
	\D{f_1f_2\gamma}{L}\,,
\end{align}
and analogous expressions for different flavours and chiralities. We will report our results for the one-loop QCD and QED shifts in terms of the amplitudes $\D{q_1q_2G}{L}$ and $\D{f_1f_2\gamma}{L}$, respectively.

\begin{table}[t]
\renewcommand{\arraystretch}{1.5}
\small
\begin{align*}
\hspace{-0.5cm}
\begin{array}[t]{|c|c|c|c}
\toprule
 \text{Operator} & \text{QCD shift} &  \text{QED shift}   \\
\midrule\midrule
\Op{V}{LR}{q_1q_3q_3q_2} &  \frac{m_{q_3}}{m_q}\D{q_1q_2G}{R} &   A_{q_3}\D{q_1q_2\gamma}{R}  \\
\cop{V}{LR}{q_1q_3q_3q_2} &  0 & N_c A_{q_3}\D{q_1q_2\gamma}{R}  \\
\Op{V}{LR}{q_1q_2q_2q_1} &  \frac{m_{q_2}}{m_q}\D{q_1q_1G}{R}+\frac{m_{q_1}}{m_q}\D{q_2q_2G}{L} &  A_{q_2}\D{q_1q_1\gamma}{R}+A_{q_1}\D{q_2q_2\gamma}{L}  \\
\cop{V}{LR}{q_1q_2q_2q_1} &  0 &  N_c\left(A_{q_2}\D{q_1q_1\gamma}{R}+A_{q_1}\D{q_2q_2\gamma}{L}\right)  \\
\Op{V}{LR}{q_1q_1q_1q_2} &  \frac{m_{q_1}}{m_q}\D{q_1q_2G}{R} & A_{q_1}\D{q_1q_2\gamma}{R}  \\
\cop{V}{LR}{q_1q_1q_1q_2} &  0 &N_c A_{q_1}\D{q_1q_2\gamma}{R}  \\
\Op{V}{LR}{q_1q_1q_1q_1} &  2\frac{m_{q_1}}{m_q}\D{q_1q_1G}{\1} & 2A_{q_1}\D{q_1q_1\gamma}{\1}  \\
\cop{V}{LR}{q_1q_1q_1q_1} &  0 &  2N_cA_{q_1}\D{q_1q_1\gamma}{\1} \\
\bottomrule
\end{array}
\end{align*}
\setlength{\belowcaptionskip}{-0.2cm}
\caption{One-loop QCD and QED shifts for VLR four-quark operators. The shifts are given in units of $(\frac{\alpha_s}{4\pi})$ and $(\frac{\alpha}{4\pi})$ for QCD and QED respectively.}
\label{tab:V4quark}
\end{table}
For the computation we will adopt the naive dimensional regularization (NDR) scheme. One complication arises however when tensor operators are considered, since closed-topology diagrams with tensor-operator insertions will include traces involving $\gamma_5$ which cannot be evaluated in NDR (see the last term of Eq.~\eqref{eq:closed}). To treat these,
we use the `t Hooft Veltman (HV) scheme, defining $D$-dimensional, four-dimensional,
and $(D - 4)$-dimensional gamma matrices $\gamma^\mu$, $\tilde{\gamma}^\mu$, and $\hat{\gamma}^\mu$,
respectively. Additionally,
\begin{equation}
	\gamma_5 = \frac{i}{4!}\epsilon^{(4)}_{\mu\nu\sigma\rho}\gamma^\mu\gamma^\nu\gamma^\sigma
	\gamma^\rho\,,
\end{equation}
where $\epsilon^{(4)}_{\mu\nu\sigma\rho}$ is the four-dimensional Levi-Civita tensor. In evaluating
diagrams with problematic traces this Levi-Civita tensor is contracted with $D$-dimensional gamma
matrices, introducing the evanescent operators

\begin{align}
	E_{q_1q_2 G}^B&=\frac{m_q}{g_s}(\overline q_1\frac{i}{2}\big( [\gamma^\mu,\hat\gamma^\nu]
  +[\hat\gamma^\mu,\gamma^\nu] \notag \\
	&-[\hat\gamma^\mu,\hat\gamma^\nu] \big)P_B T^Aq_2)G^A_{\mu\nu}
	+a\epsilon D_{q_1q_2 G}^B\,,\label{eq:evanescent_qcd} \\
	E_{f_1f_2 \gamma}^B&=\frac{m_f}{e}(\overline f_1\frac{i}{2}\big( [\gamma^\mu,\hat\gamma^\nu]
  +[\hat\gamma^\mu,\gamma^\nu] \notag \\
	&-[\hat\gamma^\mu,\hat\gamma^\nu] \big)P_B f_2)F_{\mu\nu}
  +b\epsilon D_{f_1f_2 \gamma}^B\,,
	\label{eq:evanescent_qed}
\end{align}
where we have included a non-trivial scheme-dependence parameterized by $a$ and $b$\footnote{
	Here, one may also compute such diagrams in the Larin scheme, introducing the evanescent operator
	\begin{equation}
		E'^{L/R}_{f_1 f_2 \gamma} = \frac{i m_f}{2 e}\epsilon^{(4)}_{\mu\nu\rho\sigma}\big(\bar{f}_1\sigma^{\mu\nu}P_{L/R}\,f_2\big)F^{\rho\sigma}
			\pm (1 \pm b\epsilon)D^{L/R}_{f_1 f_2 \gamma}\,,
	\end{equation}
	and analogous for the gluon operator. The resulting shifts are identical at one-loop, though differences can arise at higher orders from one-loop evanescent operator insertions.}.
The results presented depend on the particular choice of renormalization scheme. For the sake of simplicity, we choose the scheme where $a = b = 0$, which preserves the
tree-level Fierz relation, $T^{LR}_{f_1f_2f_3f_4} = 0$. However our results can, in principle, be converted to a different choice of scheme via known methods 
(see e.g. Refs~\cite{Gorbahn:2004my,Brod:2010mj}).

At the one-loop level, physical operators mix into the evanescent operators in Eqs.~\eqref{eq:evanescent_qcd} and~\eqref{eq:evanescent_qed} with $1/\epsilon$ poles. Since only the tensor operator insertions into closed-penguin topologies feature such divergent mixings, these poles are not cancelled when subtracting Fierz-conjugated insertions.

\begin{table}[t]
\renewcommand{\arraystretch}{1.5}
\small
\begin{align*}
\begin{array}[t]{|c|c|c|c}
\toprule
 \text{Operator} & \text{QCD shift} &  \text{QED shift}   \\
\midrule\midrule
\Op{V}{LR}{\ell_1q_1q_1\ell_2} &  0 &   N_cA_{q_1}\D{\ell_1\ell_2\gamma}{R}  \\
\Op{V}{LR}{q_1\ell_1\ell_1q_2} &  0 &  A_{\ell_1}\D{q_1q_2\gamma}{R}  \\
\Op{V}{LR}{\ell_1q_1q_1\ell_1} &  0 & N_c A_{q_1}\D{\ell_1\ell_1\gamma}{R}+ A_{\ell_1}\D{q_1q_1\gamma}{L}  \\
\Op{V}{LR}{\ell_1\ell_2\ell_2\ell_3} &  0 &  A_{\ell_2}\D{\ell_1\ell_3\gamma}{R}  \\
\Op{V}{LR}{\ell_1\ell_2\ell_2\ell_1} &  0 & A_{\ell_2}\D{\ell_1\ell_1\gamma}{R}+A_{\ell_1}\D{\ell_2\ell_2\gamma}{L}  \\
\Op{V}{LR}{\ell_1\ell_1\ell_1\ell_2} &  0 &  A_{\ell_1}\D{\ell_1\ell_2\gamma}{R}  \\
\Op{V}{LR}{\ell_1\ell_1\ell_1\ell_1} &  0 & 2A_{\ell_1}\D{\ell_1\ell_1\gamma}{\1}  \\
\bottomrule
\end{array}
\end{align*}
\caption{One-loop QED shifts for VLR semi-leptonic and four-lepton operators. The shifts are given in units of $(\frac{\alpha}{4\pi})$ for QED, whereas the QCD shifts all vanish.}
\label{tab:VSL4L}
\end{table}

They are, however, treated by enforcing tree-level relations $\mathcal{F}E^B_{q_1q_2G} = \mathcal{F}E^B_{q_1q_2\gamma} = 0$, consistent with the requirement that evanescent operators vanish in the limit $D\to4$. In general, these relations will also obtain loop-level shifts which become relevant at higher orders.

\vspace{-0.4cm}

\section{Results}\label{sec:res}
In this section we report the obtained shifts from dipole operators to all four-fermion operators given in Eqs~\eqref{eq:V4q}-\eqref{eq:T4l}.
We find no additional shifts to any four-fermion operators with $V^{AA}$ or $S^{AA}$ Dirac structures as well as to operators with all different fermion flavors. Therefore, we only report QCD and QED shifts for the $V^{LR}$ and $T^{LL}$ operators with at least two equal flavors. The results for the chirality-flipped operators $V^{RL}$ and $T^{RR}$ are obtained by the replacement $P_L \leftrightarrow P_R$. Results are presented in Tabs.~\ref{tab:V4quark},~\ref{tab:VSL4L} for vector operator insertions and in Tabs.~\ref{tab:T4quark}~-~\ref{tab:T4L} for tensor operator insertions. To shorten the notation we introduce the abbreviation:
\begin{equation}
	A_{f'}\equiv \frac{m_{f'}}{m_f}Q_{f'}\,.
\end{equation}
Results for ${\widetilde S}^{AA}$ and ${\widetilde T}^{AA}$ operator insertions can be obtained from Tabs.~\ref{tab:T4quark}~-~\ref{tab:T4L} in conjunction with corresponding tree-level Fierz relations. Additionally, the scheme-dependencies arising from the definition of the evanescent operators in Eqs.~\ref{eq:evanescent_qcd} and~\ref{eq:evanescent_qed} are set to zero, i.e. $a = b = 0$. Only tensor and scalar operators obtain scheme-dependent shifts, e.g.\footnote{All other scheme-dependent shifts can be obtained from these examples with proper permutations of fields in the operators and tree-level Fierz relations.}
\begin{align}
	&\expval{\widetilde{T}_{q_1q_2q_3q_3}^{AB}}_{q_1q_2G}^{(1)} = -\frac{\alpha_s}{2\pi}a \frac{m_{q_3}}{m_q}\expval{\D{q_1q_2G}{A}}_{q_1q_2G}^{(0)}\,, \\
	&\expval{T_{q_1q_3q_3q_2}^{AA} + 6\widetilde{S}_{q_1q_2q_3q_3}^{AA} - \frac{1}{2}\widetilde{T}^{AA}_{q_1q_2q_3q_3}}_{q_1q_2G}^{(1)} \nonumber\\
	&\hspace{1.5cm} = -\frac{\alpha_s}{4\pi}\Big(1 + a\Big)\frac{m_{q_3}}{m_q}\expval{\D{q_1q_2G}{A}}_{q_1q_2G}^{(0)} \,.
\end{align}
Note that operators with Dirac structure $T^{AB}$ with $A\neq B$ obtain non-vanishing scheme-dependent shifts from projections onto evanescent dipole operators and therefore should be included in the basis of operators in calculations where $a,b\neq 0$.



\begin{table}[t]
\renewcommand{\arraystretch}{1.5}
\small
\begin{align*}
\hspace{-0.75cm}
\begin{array}[t]{|c|c|c|c}
\toprule
 \text{Operator} & \text{QCD shift} &  \text{QED shift}   \\
\midrule\midrule
\Op{T}{LL}{q_1q_3q_3q_2} &  -\frac{m_{q_3}}{m_q}\D{q_1q_2G}{L} & -A_{q_3}\D{q_1q_2\gamma}{L}  \\
\Op{T}{LL}{q_1q_2q_3q_3} &  0 & \frac{N_c}{2}A_{q_3}\D{q_1q_2\gamma}{L}  \\
\Op{T}{LL}{q_1q_2q_2q_1} &  -\frac{m_{q_2}}{m_q}\D{q_1q_1G}{L}-\frac{m_{q_1}}{m_q}\D{q_2q_2\gamma}{L} & -A_{q_2}\D{q_1q_1\gamma}{L}-A_{q_1}\D{q_2q_2\gamma}{L}  \\
\Op{T}{LL}{q_1q_1q_2q_2} &  0 & \frac{N_c}{2}\left(A_{q_2}\D{q_1q_1\gamma}{L}+A_{q_1}\D{q_2q_2\gamma}{L}\right)  \\
\Op{T}{LL}{q_1q_1q_1q_2} &  -\frac{m_{q_1}}{m_q}\D{q_1q_2G}{L} & (\frac{N_c}{2}-1)A_{q_1}\D{q_1q_2\gamma}{L}  \\
\Op{T}{LL}{q_1q_1q_1q_1} &  -2\frac{m_{q_1}}{m_q}\D{q_1q_1G}{L}& (N_c-2)A_{q_1}\D{q_1q_1\gamma}{L}   \\
\bottomrule
\end{array}
\end{align*}
\caption{One-loop QCD and QED shifts for four-quark TLL operators. The shifts are given in units of $(\frac{\alpha_s}{4\pi})$ and $(\frac{\alpha}{4\pi})$ for QCD and QED respectively.}
\label{tab:T4quark}
\end{table}

\section{Applications}
As an application of the results presented in Sec.~\ref{sec:res} we discuss the simple example of computing one-loop corrections to the muonic electric dipole operator $D_{\mu\mu\gamma}^R$ in the Weak Effective theory (WET) below the electroweak scale. Matching conditions from the SM Effective Theory (SMEFT) onto the WET, known at tree-level \cite{Aebischer:2015fzz,Jenkins:2017jig} and since recently also at one-loop \cite{Dekens:2019ept}, as well as the one-loop WET renormalization group equations \cite{Jenkins:2017dyc,Aebischer:2017gaw} are traditionally given in the JMS basis introduced in \cite{Jenkins:2017jig}. This contains the semi-leptonic tensor operators of the form\footnote{We adopt the \texttt{WCxf} \cite{Aebischer:2017ugx} convention for the JMS basis.}

\begin{equation}
	O^{T,RR}_{\substack{eu\\ijkl}} = (\overline e^i \sigma_{\mu\nu} P_R e^j)(\overline u^k \sigma^{\mu\nu} P_R u^l)\,.
\label{eq:badtensor}
\end{equation}

Such operators are, however, problematic in the case where two fermion flavors on the same current are equal, since closed penguin topologies occur. These diagrams contain traces such as ${\rm Tr}\left[\gamma_\mu\gamma_\nu\gamma_\sigma\gamma_\rho\gamma_5\right]$, which are ill-defined in NDR. For this reason it is more convenient to use Fierz relations to eliminate such operators from the basis in favor of those which do not lead to problematic traces. This allows one to use the more convenient NDR scheme for the one-loop calculation.

\begin{table}[t]
\renewcommand{\arraystretch}{1.5}
\small
\begin{align*}
\hspace{-0.3cm}
\begin{array}[t]{|c|c|c|c}
\toprule
 \text{Operator} & \text{QCD shift} &  \text{QED shift}   \\
\midrule\midrule
\Op{T}{LL}{\ell_1q_1q_1\ell_2} & 0 & -N_cA_{q_1}\D{\ell_1\ell_2\gamma}{L}  \\
\Op{T}{LL}{q_1\ell_1\ell_1q_2} &  0 & -A_{\ell_1}\D{q_1q_1\gamma}{L}\\
\Op{T}{LL}{q_1\ell_1\ell_1q_1} &  0 & -N_cA_{q_1}\D{\ell_1\ell_1\gamma}{L}-A_{\ell_1}\D{q_1q_1\gamma}{L}  \\
\Op{T}{LL}{q_1q_1\ell_1\ell_2} &  0 & \frac{N_c}{2}A_{q_1}\D{\ell_1\ell_2\gamma}{L}   \\
\Op{T}{LL}{q_1q_2\ell_1\ell_1} &  0 & \frac{1}{2}A_{\ell_1}\D{q_1q_2\gamma}{L}  \\
\Op{T}{LL}{q_1q_1\ell_1\ell_1} &  0 & \frac{1}{2}\left(N_cA_{q_1}\D{\ell_1\ell_1\gamma}{L}+A_{\ell_1}\D{q_1q_1\gamma}{L}\right)  \\
\bottomrule
\end{array}
\end{align*}
\caption{One-loop QED shifts for semi-leptonic TLL operators. The shifts are given in units of $(\frac{\alpha}{4\pi})$ for QED, whereas the QCD shifts all vanish.}
\label{tab:TSL}
\end{table}

After Fierz-conjugating the tensor operator using the procedure in Refs.~\cite{Aebischer:2022aze,Chetyrkin:1997gb,Gorbahn:2004my}, we are left with the computation of only open penguin diagrams with scalar and tensor operator insertions, which can be done consistently in NDR.

This gives the following renormalized one-loop matrix element of the dipole operator:
\begin{equation}
	-i\frac{e}{16\pi^2}N_c m_\mu \sum_p A_p (\overline \mu \slashed{q}\gamma_\mu P_R\mu)\varepsilon^\mu(q)\left[1+8\log{\left(\frac{m_p^2}{\mu^2}\right)}\right]\,,
\end{equation}
with the renormalization scale $\mu$. Applying now our shifts in Tab.~\ref{tab:TSL} leads to the contribution to the muon magnetic moment
\begin{equation}
	a^{2\ell 2q}_\mu = -m_\mu^2\sum_{p}\frac{N_c A_p}{Q_\mu\pi^2}\log\Big(\frac{\mu^2}{m_p^2}\Big)\Re \left[L^{T,RR}_{\substack{eu\\\mu\mu pp}}(\mu)\right]\,,
\end{equation}
which agrees with Eq.~(3.4) of \cite{Aebischer:2021uvt}. Using the shifts, we are able to consistently combine the results of the Wilson coefficient $L^{T,RR}_{\substack{eu\\\mu\mu pp}}$, whose matching contributions are calculated in the JMS basis, with the one-loop amplitudes calculated in the Fierz-conjugated basis, where the NDR scheme can be used in the full calculation.

\section{Conclusions}

In this letter, we have expanded upon the results in Ref.~\cite{Aebischer:2022aze} by including the one-loop shifts in the four-fermion Fierz relations including QCD and QED dipole operators.

We illustrated the usefulness of our results by employing the one-loop shifts in the computation of a one-loop matrix element. This allowed us to use a simpler operator basis that avoided complications involving traces including $\gamma_5$, while still using the known matching results for the JMS basis. The one-loop Fierz transformations can also be used in matching calculations, where the model matches onto a basis which needs to be Fierz-conjugated. This is the case, for instance, in LQ models, where the matching is performed in the LQ basis and the results are then transformed into the SM basis using Fierz relations.

Another interesting application is to include Fierz-conjugation of four-fermion operators using shifts in codes such as \texttt{abc\_eft} \cite{Proceedings:2019rnh}. One could additionally extend this work to include two-loop shifts, which are relevant for two-loop dipole calculations, where both tensor and scalar operator insertions into closed-penguin topologies lead to problematic traces involving $\gamma_5$. Operator insertions with at least one different fermion flavor can be exchanged for their open-penguin counterparts using Fierz relations along with corresponding shifts.

\section{Note Added}

The present work resulted after common discussions with the authors of~\cite{Fuentes-Martin:2022vvu}. The results presented in this article are complementary to the ones in Ref.~\cite{Fuentes-Martin:2022vvu} in the sense that they are applicable in the WET below the EW scale, whereas in \cite{Fuentes-Martin:2022vvu} the SMEFT together with the full SM gauge group was considered.

\begin{table}[t]
\renewcommand{\arraystretch}{1.5}
\small
\begin{align*}
\hspace{-0.3cm}
\begin{array}[t]{|c|c|c|c}
\toprule
 \text{Operator} & \text{QCD shift} &  \text{QED shift}   \\
\midrule\midrule
\Op{T}{LL}{\ell_1\ell_2\ell_2\ell_3} & 0 & -A_{\ell_2}\D{\ell_1\ell_3\gamma}{L}  \\
\Op{T}{LL}{\ell_1\ell_2\ell_3\ell_3} &  0 & \frac{1}{2}A_{\ell_3}\D{\ell_1\ell_2\gamma}{L}\\
\Op{T}{LL}{\ell_1\ell_2\ell_2\ell_1} &  0 & -A_{\ell_2}\D{\ell_1\ell_1\gamma}{L}-A_{\ell_1}\D{\ell_2\ell_2\gamma}{L}  \\
\Op{T}{LL}{\ell_1\ell_1\ell_2\ell_2} &  0 & \frac{1}{2}\left(A_{\ell_2}\D{\ell_1\ell_1\gamma}{L}+A_{\ell_1}\D{\ell_2\ell_2\gamma}{L}\right)   \\
\Op{T}{LL}{\ell_1\ell_1\ell_1\ell_2} &  0 & -\frac{1}{2}A_{\ell_1}\D{\ell_1\ell_2\gamma}{L}  \\
\Op{T}{LL}{\ell_1\ell_1\ell_1\ell_1} &  0 & -A_{\ell_1}\D{\ell_1\ell_1\gamma}{L}  \\
\bottomrule
\end{array}
\end{align*}
\caption{One-loop QED shifts for leptonic TLL operators. The shifts are given in units of $(\frac{\alpha}{4\pi})$ for QED, whereas the QCD shifts all vanish.}
\label{tab:T4L}
\end{table}

\section*{Acknowledgements}
We thank Gino Isidori and Joachim Brod for useful discussions. This project has received funding from the European Research Council (ERC) under the European Union's Horizon 2020 research and innovation programme under grant agreement 833280 (FLAY), and by the Swiss National Science Foundation (SNF) under contract 200020\_204428.

\appendix

\section{Tree-Level Fierz Relations}

In this appendix, we collect the tree-level Fierz relations necessary for the calculation of the one-loop shifts arising from dipole operators. Using the notation of Eqs.~(\ref{eq:V4q}-\ref{eq:T4l})
with $A,B \in \{R,L\}$, we use the four-quark operator relations
\begin{equation}
	\begin{split}
		V^{AA}_{q_1q_2q_3q_4} \to & \widetilde{V}^{AA}_{q_1q_4q_3q_2}\,, \\[0.5em]
		V^{AB}_{q_1q_2q_3q_4} \to & - 2 \widetilde{S}^{BA}_{q_1q_4q_3q_2} \quad (A\neq B)\,, \\[0.5em]
		S^{AA}_{q_1q_2q_3q_4} \to & -\frac{1}{2}\widetilde{S}^{AA}_{q_1q_4q_3q_2} - \frac{1}{8}\widetilde{T}^{AA}_{q_1q_4q_3q_2}\,, \\[0.5em]
		S^{AB}_{q_1q_2q_3q_4} \to & -\frac{1}{2}\widetilde{V}^{BA}_{q_1q_4q_3q_2} \quad (A\neq B)\,, \\[0.5em]
		T^{AA}_{q_1q_2q_3q_4} \to & -6\widetilde{S}^{AA}_{q_1q_4q_3q_2} +\frac{1}{2}\widetilde{T}^{AA}_{q_1q_4q_3q_2}\,, \\[0.5em]
	\end{split}
\end{equation}
semi-leptonic operator relations
\begin{equation}
	\begin{split}
		V^{AA}_{q_1q_2\ell_1\ell_2} \to & V^{AA}_{q_1\ell_2\ell_1q_2}\,, \\[0.5em]
		V^{AB}_{q_1q_2\ell_1\ell_2} \to & - 2 S^{BA}_{q_1\ell_2\ell_1q_2} \quad (A\neq B)\,, \\[0.5em]
		S^{AA}_{q_1q_2\ell_1\ell_2} \to & -\frac{1}{2}S^{AA}_{q_1\ell_2\ell_1q_2} - \frac{1}{8}T^{AA}_{q_1\ell_2\ell_1q_2}\,, \\[0.5em]
		S^{AB}_{q_1q_2\ell_1\ell_2} \to & -\frac{1}{2}V^{BA}_{q_1\ell_2\ell_1q_2} \quad (A\neq B)\,, \\[0.5em]
		T^{AA}_{q_1q_2\ell_1\ell_2} \to & -6S^{AA}_{q_1\ell_2\ell_1q_2} +\frac{1}{2}T^{AA}_{q_1\ell_2\ell_1q_2}\,, \\[0.5em]
	\end{split}
\end{equation}
the four-lepton operator relations
\begin{equation}
	\begin{split}
		V^{AA}_{\ell_1\ell_2\ell_3\ell_4} \to & V^{AA}_{\ell_1\ell_4\ell_3\ell_2}\,, \\[0.5em]
		V^{AB}_{\ell_1\ell_2\ell_3\ell_4} \to & - 2 S^{BA}_{\ell_1\ell_4\ell_3\ell_2} \quad (A\neq B)\,, \\[0.5em]
		S^{AA}_{\ell_1\ell_2\ell_3\ell_4} \to & -\frac{1}{2}S^{AA}_{\ell_1\ell_4\ell_3\ell_2} - \frac{1}{8}T^{AA}_{\ell_1\ell_4\ell_3\ell_2}\,, \\[0.5em]
		S^{AB}_{\ell_1\ell_2\ell_3\ell_4} \to & -\frac{1}{2}V^{BA}_{\ell_1\ell_4\ell_3\ell_2} \quad (A\neq B)\,, \\[0.5em]
		T^{AA}_{\ell_1\ell_2\ell_3\ell_4} \to & -6S^{AA}_{\ell_1\ell_4\ell_3\ell_2} +\frac{1}{2}T^{AA}_{\ell_1\ell_4\ell_3\ell_2}\,, \\[0.5em]
	\end{split}
\end{equation}
as well as the tensor relation $T^{AB}_{f_1f_2f_3f_4} \to 0$ for $A~\neq~B$ which holds for four-quark, semi-leptonic, and four-lepton operators.

\bibliography{refs}
\end{document}